\RequirePackage[running]{lineno}

\documentclass[aps,prl,nofootinbib,superscriptaddress,letterpaper,amsmath,amssymb]{revtex4}

\usepackage{graphicx}  
\usepackage{dcolumn}   
\usepackage{bbm}        
\usepackage{amsmath}
\usepackage{amsfonts}
\usepackage{amssymb}   
\usepackage{feynmp}
\usepackage{epstopdf}
\usepackage{slashed}
\usepackage{multirow}
\usepackage{color}
\usepackage{float}

\usepackage{verbatim}

\def\gsim{\lower0.5ex\hbox{$\:\buildrel >\over\sim\:$}}
\def\lsim{\lower0.5ex\hbox{$\:\buildrel <\over\sim\:$}}

\let\m=\mu

\newcommand{\be}{\begin{equation}}
\newcommand{\ee}{\end{equation}}
\newcommand{\bea}{\begin{eqnarray}}
\newcommand{\eea}{\end{eqnarray}}

\newcommand{\nbox}{{\,\lower0.9pt\vbox{\hrule \hbox{\vrule height 0.2 cm
\hskip 0.2 cm \vrule height 0.2 cm}\hrule}\,}}
 \newcommand{\vev}[1]{\langle {#1} \rangle}
 \newcommand{\lagr}{\mathcal{L}}

\def\sub#1{_{\lower.25ex\hbox{$\scriptstyle#1$}}}

\newskip\zatskip \zatskip=0pt plus0pt minus0pt
\def\matth{\mathsurround=0pt}
\def\lsim{\mathrel{\mathpalette\atversim<}}
\def\gsim{\mathrel{\mathpalette\atversim>}}
\def\sigv{\ifmmode \langle\sigma v\rangle\else $\langle\sigma v\rangle$\fi}
\newskip\zatskip \zatskip=0pt plus0pt minus0pt
\def\matth{\mathsurround=0pt}
\def\lsim{\mathrel{\mathpalette\atversim<}}
\def\gsim{\mathrel{\mathpalette\atversim>}}
\def\atversim#1#2{\lower0.7ex\vbox{\baselineskip\zatskip\lineskip\zatskip
  \lineskiplimit
  0pt\ialign{$\matth#1\hfil##\hfil$\crcr#2\crcr\sim\crcr}}}

\begin{document}

\thispagestyle{empty}
\vspace*{-3.5cm}

\vspace{0.5in}

\title{Indirect Detection Constraints on the Model Space of Dark Matter Effective Theories }

\begin{center}
\begin{abstract}
Using limits on photon flux from Dwarf Spheroidal galaxies, we place bounds on the parameter space of  models in which Dark Matter annihilates into multiple final state particle pair channels.  We derive constraints on effective operator models with Dark Matter couplings to third generation fermions and to pairs of Standard Model vector bosons.  We present limits in various slices of model parameter space along with estimations of the region of maximal validity of the effective operator approach for indirect detection. We visualize our bounds for models with multiple final state annihilations by projecting parameter space constraints onto triangles, a technique familiar from collider physics; and we compare our bounds to collider limits on equivalent models.
\end{abstract}
\end{center}

\author{Linda M. Carpenter}
\affiliation{The Ohio State University, Columbus, OH}
\author{Russell Colburn}
\affiliation{The Ohio State University, Columbus, OH}
\author{Jessica Goodman}
\affiliation{The Ohio State University, Columbus, OH}

\maketitle


\section{Introduction}
In this era constraints on Dark Matter models are being synthesized from multiple experiments.   There has been much recent work in collider physics, focusing on Dark Matter (DM) models, which include both UV complete theories and Effective Field Theory (EFT) scenarios.  The same models studied in collider physics imply detectable signatures from Dark Matter annihilation in space.  Due to gauge invariance or other theoretical considerations, many of these models, both EFTs and simplified models, predict couplings between Dark Matter and multiple species of Standard Model particles. Thus, Dark Matter may be produced in many correlated final state channels at colliders, and may have multiple final state annihilation channels in space, which would contribute to total detectable photon or positron flux for satellite experiments.


In this work we explore the indirect detection bounds from Fermi-LAT dwarf spheroidal galaxies \cite{Ackermann:2013yva} on models where Dark Matter annihilates into multiple final state channels.  These bounds are among the tightest constraints on DM models.  We study several EFT models with dimensions 6  and 7 effective operators. We choose operators which lead to unsuppressed DM annihilation rates in indirect detection processes, and  which  are being simultaneously studied in  DM production processes at LHC. The dimension 6 operators we study are those which couple Dark Matter to third generation fermions pairs.  The dimension 7 operators we consider are vector boson portals where there DM couples to multiple pairs of SM gauge bosons.

The use of effective operators allows a great degree of model independence for Dark Matter studies, while capturing some of the important kinematic features of Dark Matter processes \cite{Goodman:2010yf,Goodman:2010ku,Goodman:2010qn}.  For some models which are completed by loops, EFT based calculations have so far provided the best means for study at colliders. The limits of the effective operator paradigm are becoming more clear for collider analyses.   In particular, UV completions of models with low scale messenger portals are less probe-able by colliders, and models with low scale effective operator cut-offs may not be sensible at collider energies \cite{Papucci:2014iwa,Busoni:2013lha,Busoni:2014sya,Abdallah:2015ter}.  However, we expect that EFT analyses are reliable at the scale of indirect detection where the center of mass energy of the annihilation process is the same order as the dark matter particle mass itself.  We expect that the EFT treatment is valid down to much lower scales, perhaps for mediator sectors in range of 10's of GeV, as opposed to colliders where mediators of some hundred GeV to just under a TeV  may not be appropriate.  

We will visualize the bounds we set in two slices of the total parameter space, the plane of fixed DM annihilation rate, and the plane of effect operator coefficients where the total annihilation rate varies.  For the regions of fixed annihilation rates, we will use the constraints to produce a 2-D visualization of the bounds on a triangle, a technique familiar from collider physics \cite{Anandakrishnan:2014pva}.  We also compare bounds set with the dwarf limits to those set by collider constraints, and discuss the validity limits of EFTs for both cases.

The format of this paper is as follows, in Section I we will discuss dwarf constraints on photon flux from dark matter annihilations.  In Section II we will analyze constraints on models with non-interfering final state annihilations and present triangular visualizations of parameter space.  In Section III we will analyze a popular set of vector boson portal models with interfering final state channels.  Section IV presents results along with collider constraints and discusses EFT validity.  Section V concludes.

\section{Indirect Detection from Dwarf Spheroidal Galaxies}
Dwarf spheroidal galaxies provide some of the tightest constraints on photon flux originating from dark matter annihilation as they are believed to contain a substantial dark matter component \cite{Mateo:1998wg,McConnachie:2012vd}.   This combined with their low astrophysical background makes them a good laboratory to search for dark matter.  As no significant excess in the photon spectrum has been observed from dwarf data, we use upper bounds on photon flux obtained from Fermi-LAT data \cite{Ackermann:2013yva} to place constraints on dark matter mass and couplings in the scenarios discussed in the introduction.

The photon flux $(\text{photons cm}^{-2} \text{ s}^{-1})$ at the earth expected from annihilation of dark matter in the area of interest is given by
\be
\Phi_{\gamma}=\frac{1}{4\pi}
\sum_{\substack{f}}
\frac{\vev{\sigma v}_{f}}{2m_{\chi}^{2}}\int_{E_{\text{min}}}^{E_{\text{max}}}\left(\frac{dN_{\gamma}}{dE_{\gamma}}\right)_{f}dE_{\gamma} J.
\label{eq:gflux}
\ee
Where the J-factor $(\text{GeV}^2\text{cm}^{-5})$ is the line of sight integral of the dark matter density $\rho$, integrated over a solid angle, $\Delta\Omega$
\be
J=\int_{\Delta \Omega}\int_{l.o.s}\rho^{2}(\bold{r})dl d\Omega^{\prime}.
\ee
The additional terms in Eq. \ref{eq:gflux} are dependent on the particle properties of the dark matter.  Here, $dN_f/dE$ is the differential photon energy spectrum per annihilation to final state f, $\vev{\sigma v}_f$ is the thermally averaged DM annihilation cross section and $m_\chi$ is the dark matter mass.

The Fermi-LAT collaboration has presented dark matter constraints from observation of 25 dwarf spheroidal galaxies \cite{Ackermann:2013yva}.  Of these 25 Milky Way satellites, 15 were used to place constraints on annihilation cross sections of dark matter particles with masses between 2GeV and 10TeV into various standard model channels.  These annihilation constraints were derived assuming that the dark matter annihilates primarily to a single standard model channel.  This is not realistic when considering most  UV completions of Dark Matter models, even ones that are quite simple.  Here we discuss obtaining bounds on dark matter mass and annihilation cross sections assuming annihilation into multiple channels.

\begin{figure}[tbh-]
\centering
\includegraphics[scale=0.7]{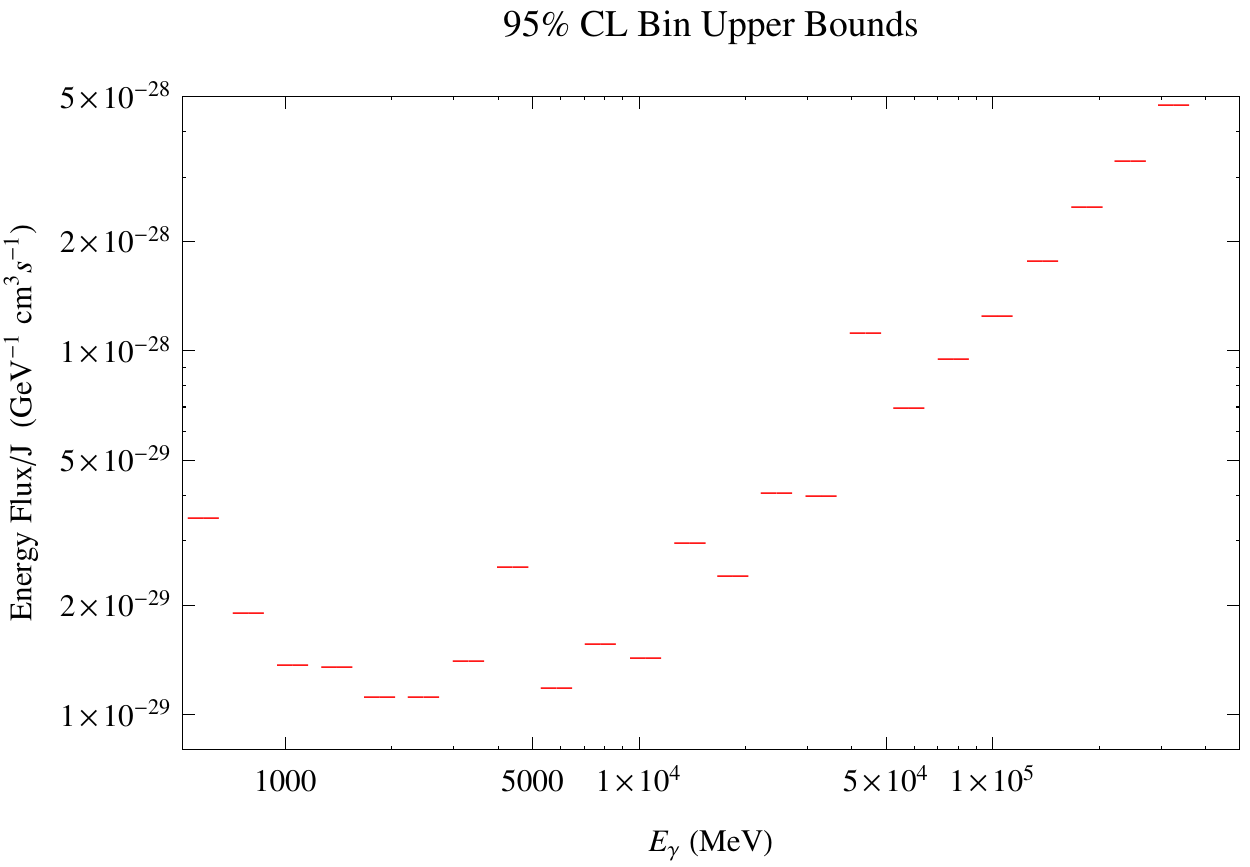}
\caption{Combined $95\%$ upper bounds on photon energy flux scaled by J-factor from 6 of largest J-factor dwarf spheroidal galaxies.}
\label{fig:CombinedDLimits}
\end{figure}

The bin by bin integrated $\gamma$-ray energy flux upper limits at $95\%$ CL for each of the 25 dwarf galaxies was presented in \cite{Ackermann:2013yva}.  The J-factor for dwarfs is determined through dynamical modeling of their star densities and velocity dispersion profiles; we use J-factors presented in \cite{Ackermann:2013yva}.  For a conservative limit we perform a very simple combination of six dwarfs with the highest J-factors.   The limit in each bin is obtained from  the dwarf which gives the strongest $95\%$ CL upper limit on $\gamma$-ray flux scaled by the J-factor.  The combined limits are shown in Fig. \ref{fig:CombinedDLimits}.

We note that the Fermi-LAT collaboration has obtained its own combined dwarf limits on photon flux from DM annihilation.  However, these limits assumed  DM annihilation into one and only one visible channel at a time.  Since our goal in this work is to analyze models that may have multiple final state annihilations we have forgone use of Fermi's combination, though in future analyses,  stronger bounds than ours may be obtained in each bin by "stacking" the total signal from all dwarves and subtracting a summed astrophysical background.

We calculate the expected $\vev{\sigma v}_{f}$ for each kinematically accessible annihilation channel, and obtain the differential gamma ray spectra for annihilation channels from the Mathematica code PPPC 4 DM ID \cite{Cirelli:2010xx,Ciafaloni:2010ti}. As an example, we show the spectrum for several relevant SM pair annihilation channels for a $100$GeV mass dark matter particle in Fig. \ref{fig:DMSpectrum100} below.  Here, the dimensionless parameter x is the gamma ray energy scaled by the dark matter mass.
\begin{figure}[tbh-]
\centering
\includegraphics[scale=0.7]{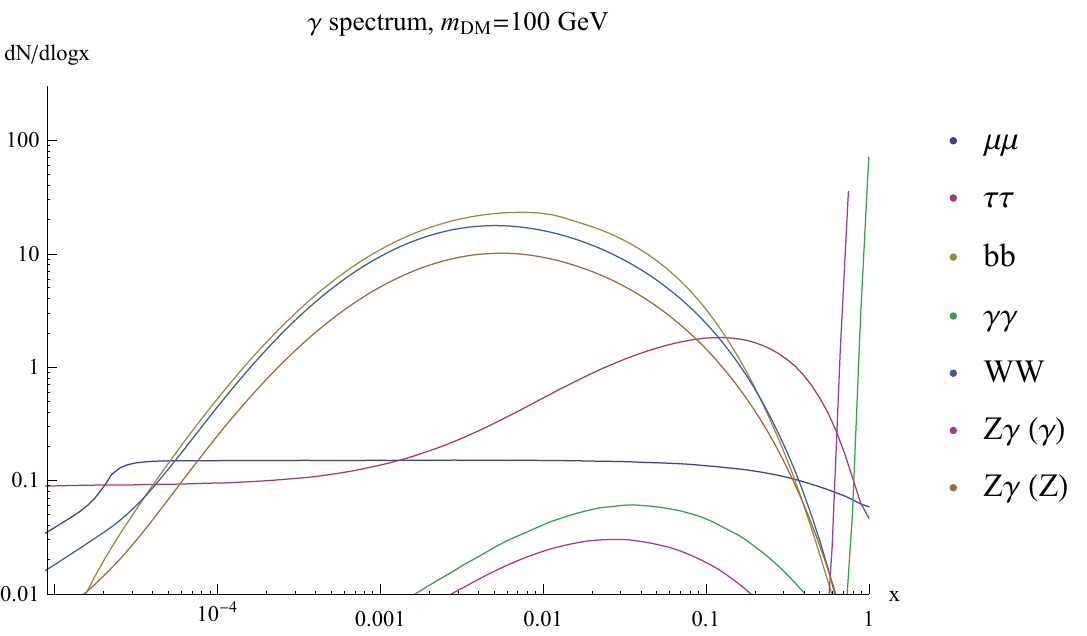}
\caption{Gamma ray spectrum from DM annihilation into various channels for a $100$GeV DM fermion.}
\label{fig:DMSpectrum100}
\end{figure}
Analysis of many models requires calculation of differential gamma ray spectra from particles produced in non-symmetric pairs, for example, in the analysis of models of section IV we have calculated the differential spectrum of the Z$\gamma$  final state.   Limits are obtained by binned comparison of total expected flux from all DM annihilations to our combined upper bounds.


\section{Models with Independent Annihilation Channels}
We will now discuss the method of constraining the parameter space of models with multiple independent annihilation channels.  We will first assume an effective Lagrangian which is the sum of several independent operators, each of which couples Dark Matter to \textbf{one and only one} pair of SM particles.  Therefore the Lagrangian has the form
\be
\mathcal{L}_{\text{f}} = \Sigma_i\mathcal{O}_{i}= \Sigma_i \frac{\kappa}{\Lambda_i^n} \chi \chi X_i X_i.
\label{eq:LGeneral}
\ee
Here the $\chi$ is the Dark Matter, and X is some Standard Model particle with particle index i. Any specific operator will be gauge and Lorentz invariant and will have coefficient $\kappa/\Lambda_i^n$, where the effective cut-off $\Lambda$ appears with the appropriate power to make the operator dimension 4.

The full parameter space of the EFT consists of the Dark Matter mass $m_{\chi}$ and the $i$ operator coefficients  $\kappa/\Lambda_i^n$.  Each point in this parameter space specifies a total DM annihilation rate, the specific ratios of the $i$ final state annihilation channels, and the resultant $\gamma$-ray flux. There are various slices of parameter space which can  be studied. The first one we will consider is slices of the parameter space on which the total DM annihilation rate is held constant.  Below we will show that along planes of parameter space with fixed annihilation rate, we will rule out masses and effective cut-offs below  certain scales.

\subsection{Fixed Annihilation Rate}

The total annihilation rate, $\vev{\sigma v}_{\text{tot}}$, is simply a linear sum of the thermally averaged annihilation cross sections $\vev{\sigma v}_{\mathcal{O}_i}$ to particle $X_i$ due to operator $\mathcal{O}_i$,
\begin{equation}
\vev{\sigma v}_{\text{tot}}=N\vev{\sigma v}_{\text{Th}}= \vev{{\sigma v}}_{\mathcal{O}_1}+\vev{{\sigma v}}_{\mathcal{O}_2}+ \cdots
\end{equation}

We will first fix the desired total annihilation rate.  This rate may be anything we like, for simplicity we will consider it some numerical factor times the thermal annihilation rate $N \vev{\sigma v}_{\text{Th}}$. This constraint drops us 1 dimension in parameter space; certain coefficient values $\kappa/\Lambda_i^n$ will satisfy the constraint for any specific DM mass. Having fixed the total annihilations rate, we may then determine the limits of the operator coefficients which saturate the Fermi-LAT photon-flux bounds for any given Dark Matter mass.

A natural choice for the total annihilation rate is the thermal rate.  However, in the spirit of model independence, we will show Fermi-bounds on models with various annihilation rates which will correspond to models with various non-thermal histories.  The complete theory will have to account for this history, as well as the presence (and absence) of specific operator coefficients.  We note that models where the total \emph{visible} annihilation rate is below the thermal rate can easily be saved from the prospect of over-closure by the addition of invisible DM annihilation channels which restore the total annihilation rate to thermal.  We discuss such models in the next subsection.

The rate constraint confines us to a hyper-surface in the full parameter space, however we may greatly simplify the form of this constraint by dividing out by the total rate $\vev{\sigma v}_{\text{tot}}$ .  We define the fractional annihilation rate as in \cite{Calore:2014nla} to be  $R_i=\vev{\sigma v}_{i}/\vev{\sigma v}_{\text{tot}}$, giving a constraint which is linear in the partial rates $R_i$.
\be
R_{1}+R_{2}+R_{3}+ \cdots =1.
\label{eq:constraint1}
\ee

\noindent The natural visualization for this parameter space of models with 3,4,5 annihilation channels etc. are a  triangle, tetrahedron, 4-simplex, etc.


As an example and for clarity of discussion we will analyze a simple model where DM may annihilate to multiple independent channels.  We choose an EFT model where dark matter couples to the third generation of fermions via  dimension 6 operators;
\be
\mathcal{L}_{\text{f}} =\frac{\kappa_t}{\Lambda_t^2}\chi \Gamma \overline{\chi} t \Gamma \overline{t}+\frac{\kappa_b}{\Lambda_b^2}\chi \Gamma \overline{\chi} b \Gamma \overline{b} + \frac{\kappa_{\tau}}{\Lambda_{\tau}^2}\chi  \Gamma \overline{\chi} \tau \Gamma \overline{\tau}+  \frac{\kappa_{\nu}}{\Lambda_{\nu}^2}\chi \Gamma \overline{\chi} \nu  \Gamma \overline{\nu}.
\label{eq:LFermions}
\ee

\begin{figure}[h]
\begin{center}
	    \begin{fmffile}{sigmav5}
	        \begin{fmfgraph*}(25,25)
                   \fmfleft{i1,i2}
                   \fmfright{o1,o2}
                   \fmfblob{.5cm}{v1}
                   \fmflabel{$\chi$}{i1}
                   \fmf{fermion}{i1,v1}
                   \fmflabel{$\bar{f}$}{o2}
                   \fmf{fermion}{v1,o2}
                   \fmflabel{$f$}{o1}
                   \fmf{fermion}{o1,v1}
                   \fmflabel{$\bar\chi$}{i2}
                   \fmf{fermion}{v1,i2}
	        \end{fmfgraph*}
	    \end{fmffile}
\end{center}
\caption{Dark matter annihilation into pairs of fermions.}
\label{fig:DMToFermions}
\end{figure}
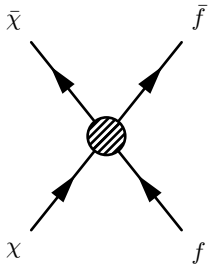

Here $\Gamma$ specifies the whether the fermionic currents are scalar, pseudo-scalar, vector etc. Under the conventions in \cite{Goodman:2010ku} they correspond to operators D1, D2, etc.  Annihilation rates will depend greatly on the Lorentz structure of the operator with operators leading to velocity and/or helicity suppression of the rate \cite{Profumo:2013hqa}.  The strongest constraints from photon-flux measurements will apply to models with operators corresponding to unsuppressed DM annihilation rates.  We have here included an annihilation into an invisible channel, specifically neutrinos.  In setting constraints for fractional annihilation rates, we may consider this channel a stand-in for annihilation into any kinematically accessible invisible channel.  The option of an invisible channel allows a total visible annihilation rate below the thermal rate while still avoiding over-closure of the universe.  Recent work, for example \cite{Falkowski:2009yz}, has shown that these annihilations may be significant in some scenarios.  The relevant annihilation diagram are shown in Fig. \ref{fig:DMToFermions}.
\vspace{0.5cm}

Let us first assume that the dark matter is light and that annihilation into top pairs in kinematically forbidden.  Here, the total annihilation rate is the sum of the thermally averaged annihilation cross section into b's, $\tau$'s, and the invisible channel,
\begin{equation}
\vev{\sigma v}_{\text{tot}}=\vev{\sigma v}_{\text{b}}+\vev{\sigma v}_{\tau}+\vev{\sigma v}_{\nu}.
\end{equation}
The crucial point here is that each operator in Eq. \ref{eq:LFermions} corresponds to only one channel, thus the constraint above has the form $\propto a \left(\kappa_b/\Lambda_b^2 \right)^2+b\left(\kappa_{\tau}/\Lambda_{\tau}^2 \right)^2+c\left(\kappa_{\nu}/\Lambda_{\nu}^2 \right)^2$ (where a, b, and c depend on dark matter mass and final state particle mass) with no terms of the form $\kappa_i \kappa_j$ for $i\ne j$.  Therefore, a triangle seems the natural choice for visualization of constraints where the sides of the triangle correspond to the $R_b,R_{\tau},R_{\nu}$ axis.

We see from Eq. \ref{eq:gflux} that once the annihilation rates $\vev{\sigma v}_{\text{f}}$ are fixed, upper bound on $\gamma$-ray flux corresponds to a lower bound on dark matter mass.  To determine this mass bound we set the total annihilation cross section to a multiple of the thermal relic, $\vev{\sigma v}_{\text{tot}}=N\vev{\sigma v}_{\text{thm}}$, then for a given point in $(R_b,R_{\tau},R_{\nu})$ space find the lowest mass which gives the maximum allowed $\gamma$-ray flux determined by Fig. \ref{fig:CombinedDLimits}.  The mass limits depend on the total annihilation rate which we have fixed; as we increase the total annihilation rate, the mass bounds become stronger.

\begin{figure}[H]
\centering
\includegraphics[scale=0.4]{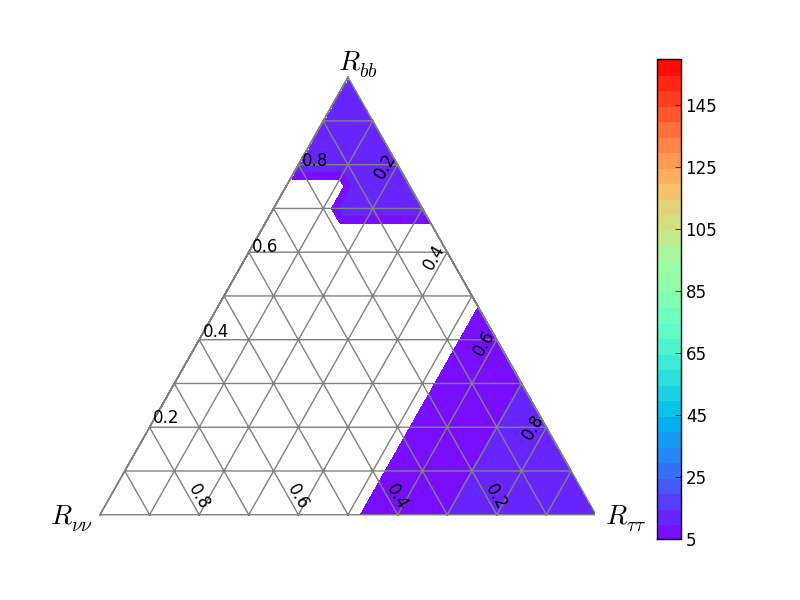}
\includegraphics[scale=0.4]{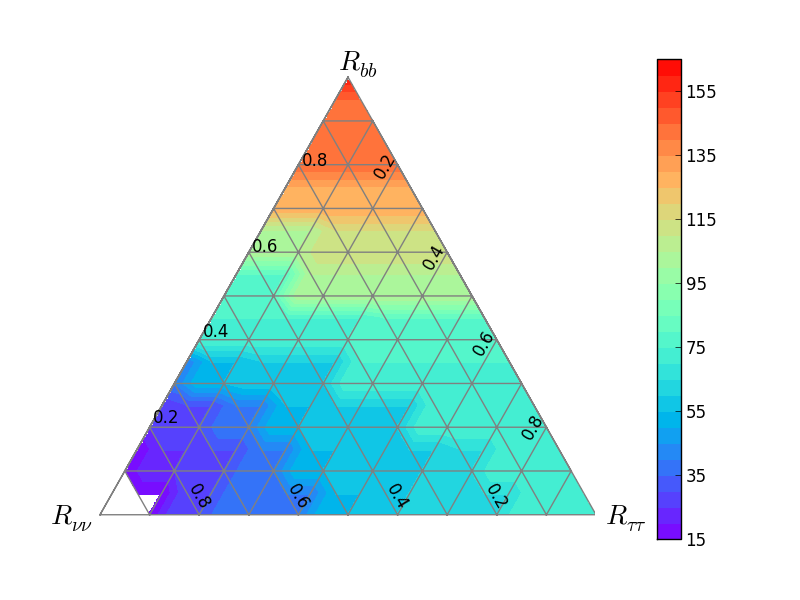}
\caption{Lower bound on dark matter mass in GeV for annihilations into b quarks, $\tau$'s, and invisible particles for $\vev{\sigma v}_{\text{tot}} = \vev{\sigma v}_{\text{Th}}$ on left and $\vev{\sigma v}_{\text{tot}} =10\vev{\sigma v}_{\text{Th}}$ on right. }
\label{fig:trel}
\end{figure}

The results are given in the Fig. \ref{fig:trel} for two total annihilation rates, the thermal rate and 10 times thermal rate.  The blank spaces in the triangle correspond to a lower bound of $0$ GeV or no lower bound.  We see that the strongest bounds come from annihilation into b's while the weakest bounds come from annihilation into invisible particles as expected.

Here the points that form the triangle each correspond to a unique ratio of the partial annihilation rates $R_i$.  At each vertex of the triangle, a single  $R_i= 1$, that is, one channel saturates that total annihilation rate.  Along the edges of the triangle only two annihilation channels contribute to the total annihilation rate, with the third partial rate set to 0.  The form of Eqn. \ref{eq:LFermions} was satisfactory to derive a lower DM mass bound for a fixed total annihilation rate.  We will now pick a specific Lorentz structure for our operators and calculate bounds on the effective operator coefficients.  As an example, we will choose to constrain the vector current operators D5, whose annihilation rate is neither velocity nor chirality suppressed.
\be
\mathcal{L}_{\text{f}} =\frac{\kappa_f}{\Lambda_f^2}\chi \gamma^{\mu} \overline{\chi} f \gamma_{\mu} \overline{f}
\ee

Each mass that satisfies the rate constraint will correspond to specific values of the effective operator coefficient.  With total annihilation rate fixed, points of lower mass will require lower effective cut-offs to satisfy the bound.  We also plot on the triangle the lowest effective cut-off (largest operator coefficient) that satisfies the bound of total flux.  Below we show the lower limits operator coefficients $ \Lambda_i^{*} = \sqrt{\kappa_i}/\Lambda_i$ which saturate the bounds on photon flux.

\begin{figure}[H]
\centering
\includegraphics[scale=0.4]{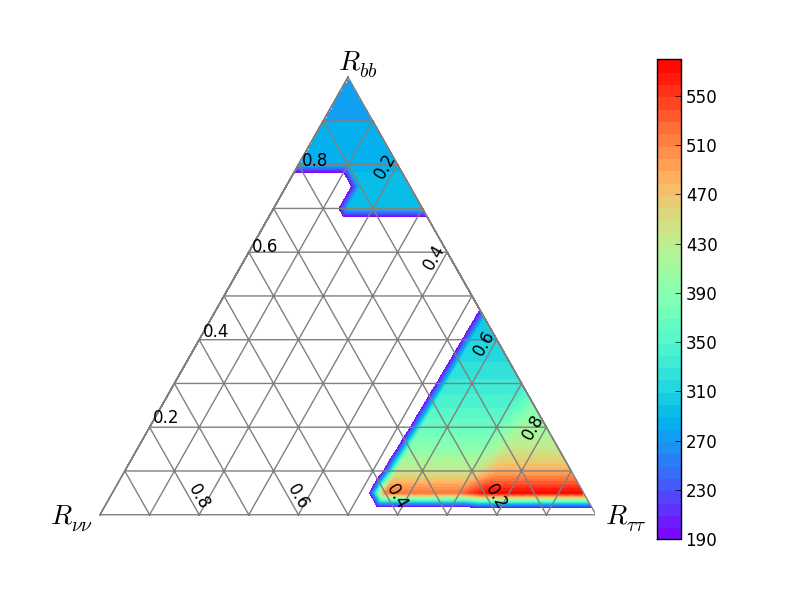}
\includegraphics[scale=0.4]{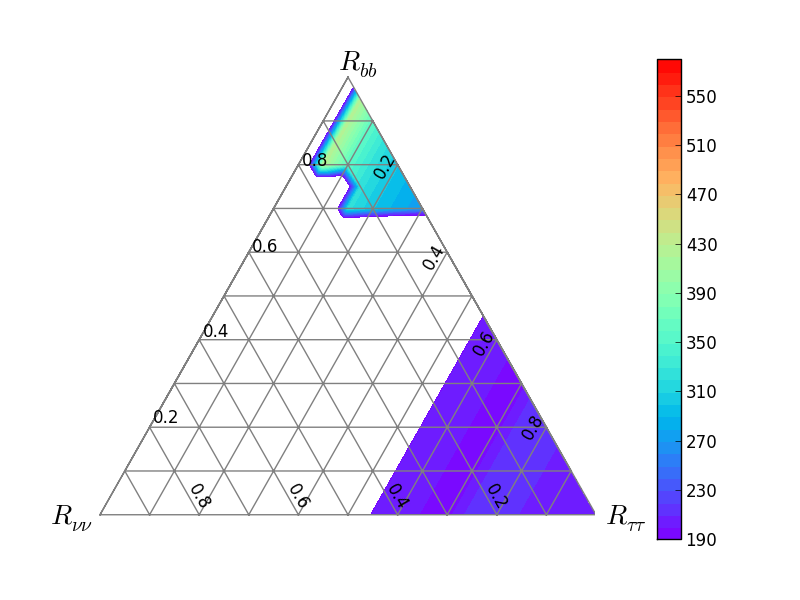}
\caption{Lower bound on operator coefficients $\Lambda_b^{*}$ and $\Lambda_{\tau}^{*}$ for  $\vev{\sigma v}_{\text{tot}} = \vev{\sigma v}_{\text{Th}}$ on left and $\vev{\sigma v}_{\text{tot}} =10\vev{\sigma v}_{\text{Th}}$ on right.}
\label{fig:LambdaLimitsTrel}
\end{figure}

For the thermal total annihilation rate, we see that DM annihilating into mostly b's or $\tau 's$, must be above 5-15 GeV in mass while the effective operator  cut-off scales may not be smaller than a few hundred GeV.  We point out that in our exclusion region, the effective operator paradigm appears to be justified as the effective cut-offs are well above the center of mass energy for the annihilation process.  We show similar triangles for a total annihilation  rate of 10 times the thermal rate in the appendix.

\subsection{Constraints for 4 Independent Channels}

\begin{figure}[h]
\centering
\includegraphics[trim={0 10cm 0 5cm},clip,scale=0.4]{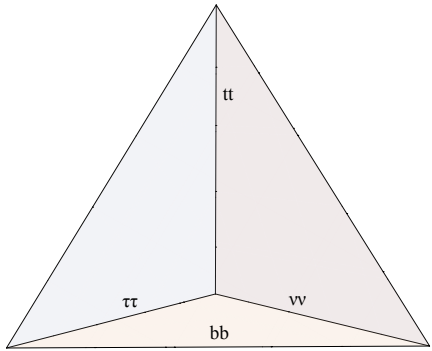}
\caption{Axis labeling for tetrahedron presentation of DM mass constraints when a fourth annihilation channel is allowed.  The base corresponds to the top left of Fig. \ref{fig:30And70TopTriangle}.}
\label{fig:blanktetrahedron}
\end{figure}

Next we consider the case of allowing more than three annihilation channels. We now allow the dark matter mass to be larger than the top mass opening up an additional annihilation channel.  One can visualize the mass constraints on a tetrahedron where the ``vertical" axis is $R_t$, see Fig. \ref{fig:blanktetrahedron}.  Each slice of the tetrahedron would correspond to a flat triangle with constraint equations given by
\begin{eqnarray}
\vev{\sigma v}_{\text{tot}}-\vev{\sigma v}_{\text{t}}&=&\vev{\sigma v}_{\text{b}}+\vev{\sigma v}_{\tau}+\vev{\sigma v}_{\nu} \nonumber \\
\left(1-R_{t}\right)&=&R_{b}+R_{\tau}+R_{\nu}.
\label{eq:constraint2}
\end{eqnarray}
\noindent

\begin{figure}[H]
\centering
\includegraphics[trim={2cm 7cm 2cm 5cm},clip,scale=0.5]{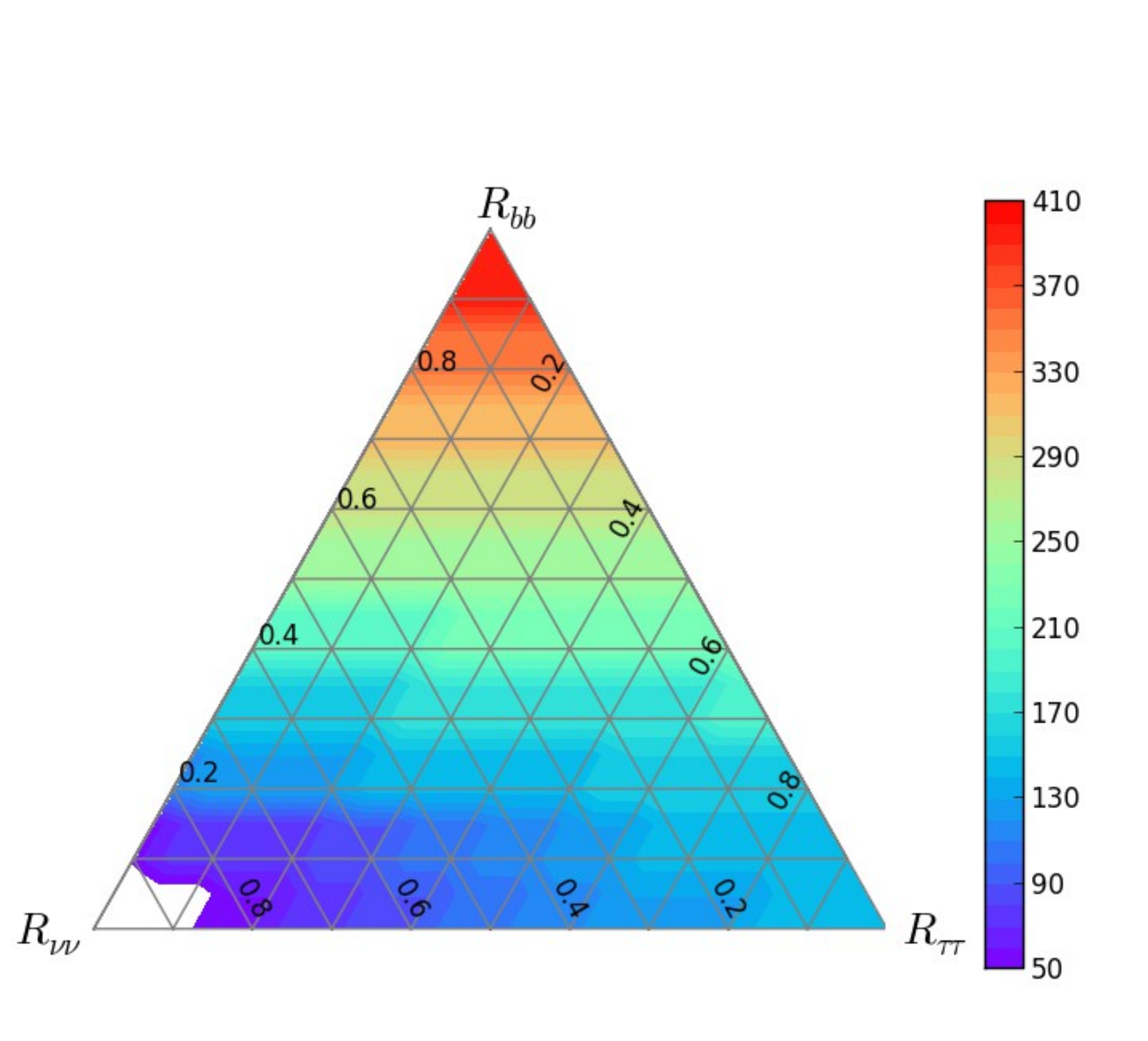}
\includegraphics[trim={2cm 7cm 2cm 5cm},clip,scale=0.5]{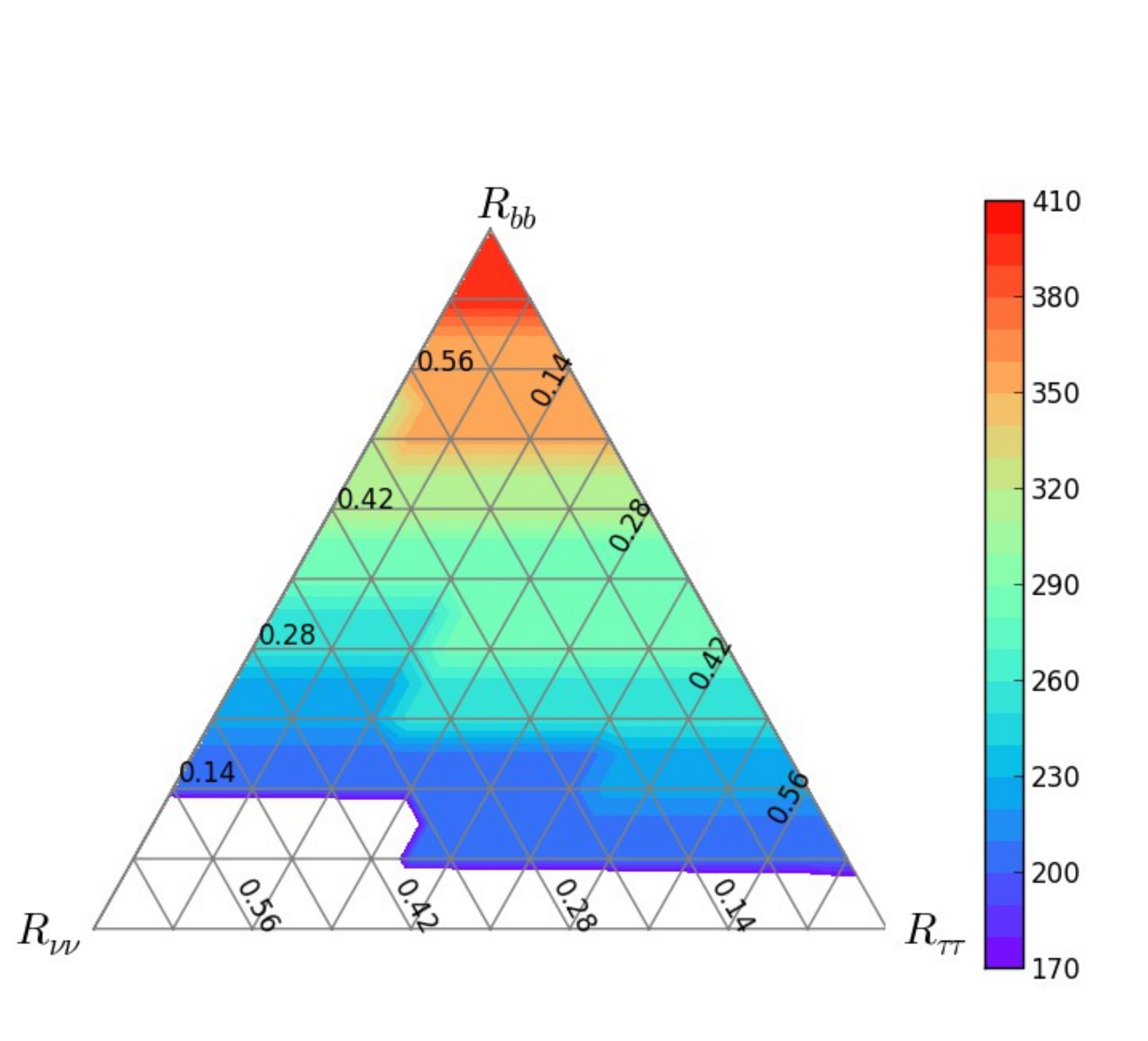}
\includegraphics[trim={2cm 7cm 2cm 5cm},clip,scale=0.5]{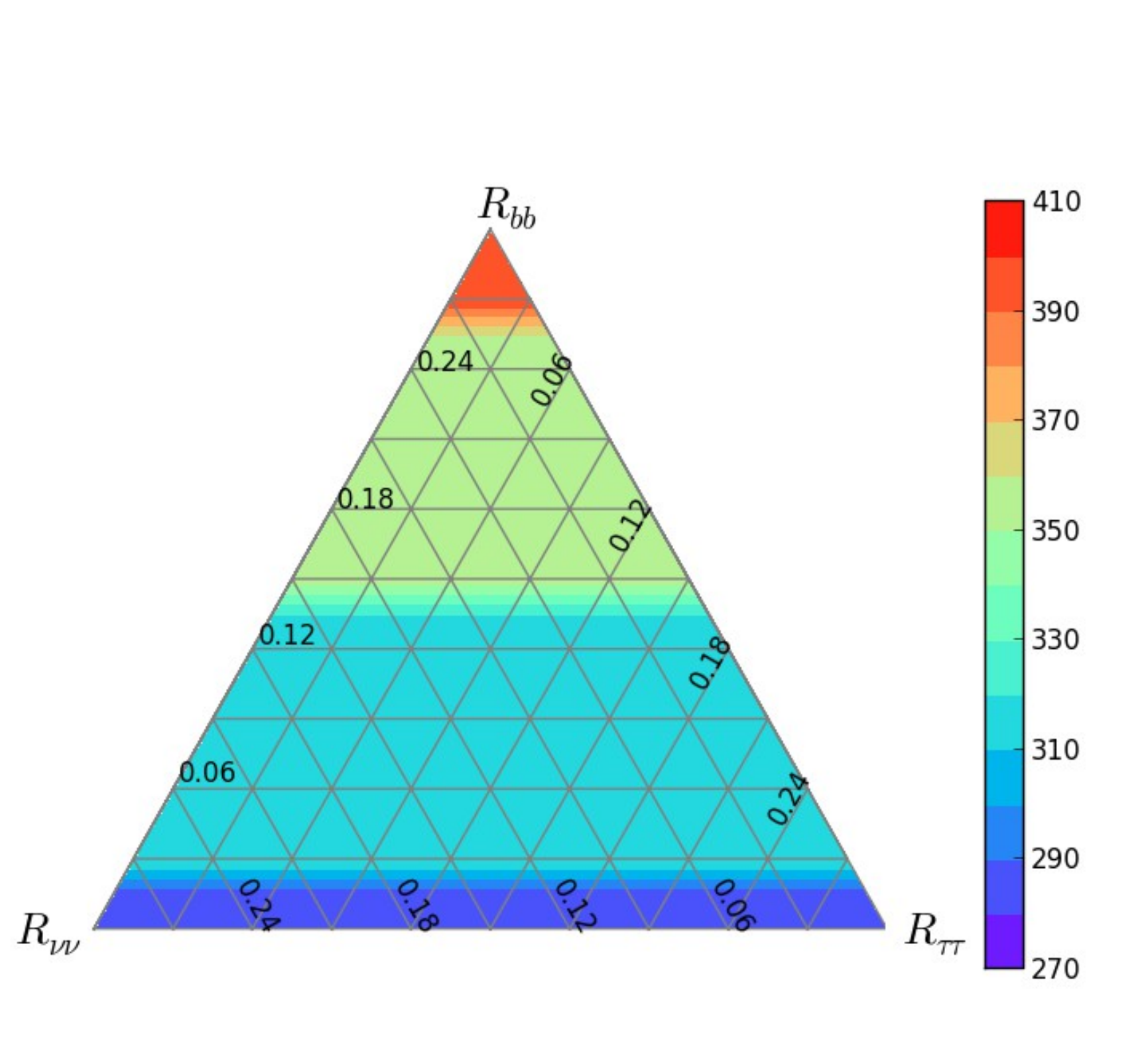}
\caption{The contours correspond to lower bounds (in GeV) on DM mass from annihilations into pairs of fermions $(tt,bb,\tau\tau,\nu\nu)$.  The total annihilation rate is set to 33 times the thermal relic for all three plots.  The left image corresponds to $R_t=0$,  the right to $R_t=.3$ and the bottom to $R_t=.7$. }
\label{fig:30And70TopTriangle}
\end{figure}

\noindent In particular, the base of the tetrahedron would correspond to $R_t=0$ and the constraints shown in the top left of Fig. \ref{fig:30And70TopTriangle}.  Below we show lower bounds on dark matter mass for triangle slices where the fractional annihilation rate to tops is $R_t=0, .3 \text{ and, }.7$.  The blank areas in the triangles of Fig. \ref{fig:30And70TopTriangle} correspond to the region of parameter space where the kinematic bound is stronger then that obtained from indirect searches.

For models with each operator corresponding to independent annihilation channels, this procedure may be iterated for any number of final state annihilation channels. In the next section we will discuss bounds on the parameter space of models where there is not a one to one correspondence between operators and annihilation channels.  We first discuss the bounds on parameter space visualized on surfaces of fixed total annihilation rate.  We will then discuss the bounds on surfaces of fixed operator coefficients.

\section{Constraints in Models with Interfering Channels}

\begin{figure}[h]
\begin{center}
	    \begin{fmffile}{sigmav}
	        \begin{fmfgraph*}(25,25)
                   \fmfleft{i1,i2}
                   \fmfright{o1,o2}
                   \fmfblob{.5cm}{v1}
                   \fmflabel{$\chi$}{i1}
                   \fmf{fermion}{i1,v1}
                   \fmflabel{$\gamma$}{o2}
                   \fmf{photon}{v1,o2}
                   \fmflabel{$\gamma$}{o1}
                   \fmf{photon}{o1,v1}
                   \fmflabel{$\bar\chi$}{i2}
                   \fmf{fermion}{v1,i2}
	        \end{fmfgraph*}
	    \end{fmffile}  \hspace{0.5cm}
	    \begin{fmffile}{sigmav1}
	        \begin{fmfgraph*}(25,25)
                   \fmfleft{i1,i2}
                   \fmfright{o1,o2}
                   \fmfblob{.5cm}{v1}
                   \fmflabel{$\chi$}{i1}
                   \fmf{fermion}{i1,v1}
                   \fmflabel{$\gamma$}{o2}
                   \fmf{photon}{v1,o2}
                   \fmflabel{Z}{o1}
                   \fmf{photon}{o1,v1}
                   \fmflabel{$\bar\chi$}{i2}
                   \fmf{fermion}{v1,i2}
	        \end{fmfgraph*}
	    \end{fmffile}
\hspace{0.5cm}
	    \begin{fmffile}{sigmav2}
	        \begin{fmfgraph*}(25,25)
                   \fmfleft{i1,i2}
                   \fmfright{o1,o2}
                   \fmfblob{.5cm}{v1}
                   \fmflabel{$\chi$}{i1}
                   \fmf{fermion}{i1,v1}
                   \fmflabel{Z}{o2}
                   \fmf{photon}{v1,o2}
                   \fmflabel{Z}{o1}
                   \fmf{photon}{o1,v1}
                   \fmflabel{$\bar\chi$}{i2}
                   \fmf{fermion}{v1,i2}
	        \end{fmfgraph*}
	    \end{fmffile}
	    \hspace{0.5cm}
	    \begin{fmffile}{sigmav3}
	        \begin{fmfgraph*}(25,25)
                   \fmfleft{i1,i2}
                   \fmfright{o1,o2}
                   \fmfblob{.5cm}{v1}
                   \fmflabel{$\chi$}{i1}
                   \fmf{fermion}{i1,v1}
                   \fmflabel{$W^+$}{o2}
                   \fmf{photon}{v1,o2}
                   \fmflabel{$W^-$}{o1}
                   \fmf{photon}{o1,v1}
                   \fmflabel{$\bar\chi$}{i2}
                   \fmf{fermion}{v1,i2}
	        \end{fmfgraph*}
	    \end{fmffile}
	    \hspace{0.5cm}
	    \begin{fmffile}{sigmav4}
	        \begin{fmfgraph*}(25,25)
                   \fmfleft{i1,i2}
                   \fmfright{o1,o2}
                   \fmfblob{.5cm}{v1}
                   \fmflabel{$\chi$}{i1}
                   \fmf{fermion}{i1,v1}
                   \fmflabel{g}{o2}
                   \fmf{gluon}{v1,o2}
                   \fmflabel{g}{o1}
                   \fmf{gluon}{o1,v1}
                   \fmflabel{$\bar\chi$}{i2}
                   \fmf{fermion}{v1,i2}
	        \end{fmfgraph*}
	    \end{fmffile}
\end{center}
\caption{Dark matter annihilation into gauge bosons.}
\label{fig:DMToGauge}
\end{figure}
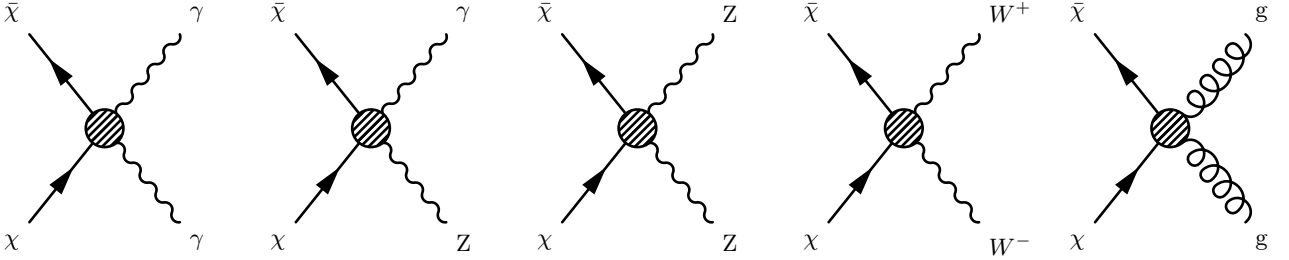

Many examples of models exist in which a single parameter or operator coefficient in the Lagrangian demands Dark Matter annihilation to multiple final state channels.  Further there may be multiple operators or parameters in the Lagrangian which contribute to annihilations into the same channel.  We will now analyze a model in which both of the previous statements are true.  We consider a popular set of effective operator models for collider and indirect studies (see, for example \cite{Chen:2013gya,Rajaraman:2012fu,Weiner:2012gm,Crivellin:2015wva,Nelson:2013pqa}), the vector boson portal models .  We will consider a set of dimension 7 operators where a pseudo-scalar Dark Matter current couples to the SM field strength tensors.
\begin{equation}
\lagr=\frac{\kappa_{1}}{\Lambda^{3}}\bar{\chi}\gamma^{5}\chi B_{\mu\nu}B^{\mu\nu}
+\frac{\kappa_{2}}{\Lambda^{3}}\bar{\chi}\gamma^{5}\chi W^{i}_{\mu\nu}W_{i}^{\mu\nu}
+\frac{\kappa_{3}}{\Lambda^{3}}\bar{\chi}\gamma^{5}\chi G^{a}_{\mu\nu}G_{a}^{\mu\nu}
\label{eq:gaugelagr}
\end{equation}
These operators are not velocity suppressed and  they may naturally arise if the dark matter  couple to the SM through loops of heavy messengers which carry SM charge \cite{Weiner:2012gm}.  Unlike the previous model discussed, gauge invariance ensures that two of the above operators contribute to multiple annihilation channels.  These two operators also have interfering contributions to the same annihilation channels.  The coefficients $\kappa_{1}/\Lambda^3$ and $\kappa_{2}/\Lambda^3$ control the coupling of dark matter to four pairs of electroweak gauge boson final states: $ZZ$, $Z\gamma$, $\gamma \gamma$, and $W^{+}W^{-}$, see the relevant diagrams in Fig. \ref{fig:DMToGauge}.
\vspace{0.5cm}
The thermally averaged annihilation cross section to di-boson final state is given in Eq. \ref{eq:xsecs} below.
\begin{equation}
\begin{split}
\langle\sigma v_{rel}\rangle_{WW}&=\frac{\kappa_{2}^{2}}{4 \pi \Lambda^{6}}\sqrt{1-\frac{m_{W}^{2}}{m_{\chi}^{2}}}(16m_{\chi}^{4}-16m_{W}^{2}m_{\chi}^{2}+6m_{W}^{4})\\
\langle\sigma v_{rel}\rangle_{ZZ}&=\frac{(\kappa_{1}s_{w}^{2}+\kappa_{2}c_{w}^{2})^{2}}{8 \pi \Lambda^{6}}\sqrt{1-\frac{m_{Z}^{2}}{m_{\chi}^{2}}}(16m_{\chi}^{4}-16m_{Z}^{2}m_{\chi}^{2}+6m_{Z}^{4})\\
\langle\sigma v_{rel}\rangle_{Z\gamma}&=\frac{s_{w}^{2}c_{w}^{2}(\kappa_{2}-\kappa_{1})^{2}}{ 16\pi m_{\chi}^{2}\Lambda^{6}}(4m_{\chi}^{2}-m_{Z}^{2})^{3}\\
\langle\sigma v_{rel}\rangle_{\gamma\gamma}&=\frac{4(\kappa_{1}c_{w}^{2}+\kappa_{2}s_{w}^{2})^{2}}{2\pi\Lambda^{6}}m_{\chi}^{4}\\
\langle\sigma v_{rel}\rangle_{gg}&=\frac{16\kappa_{3}^{2}}{\pi\Lambda^{6}}m_{\chi}^{4}\\
\end{split}
\label{eq:xsecs}
\end{equation}

We now explore the bounds on the parameter space of this model, the space consisting of the dark matter mass $m_{\chi}$ and the three operator coefficients. For any given point in parameter space, the total photon flux due to annihilations results from the sum the annihilations into each of the five di-boson channels.  We see that the operator with coefficient $\kappa_3/\Lambda^3$ controls the annihilation to gluons and only gluons, and therefore factorizes in the total annihilation rate as per the discussion before.  However the total rate to other di-boson pairs depends on two operator coefficients.  In this case, the total annihilation rate has the form $(k_i=\kappa_i/\Lambda^3)$

\be
(ak_1^2 +bk_2^2)^2+ck_3^2=\vev{\sigma v}_{tot}
\ee

We may again study the system by fixing the total annihilation rate as in the previous section.  Unlike the case in the previous section, the rate constraint is not linear; however for each specific dark matter mass, the couplings which saturate the annihilation rate are still constrained to lie on a 2-dimensional surface. This surface  is a triangular section of an ellipsoid.  The radii of the ellipsoid vary as the total annihilation rate is changed.  We can again present dwarf constrains on the space of operator coefficients using a simple visualization, now on an ellipsoidal surface rather than a triangle. In Fig 9, we show the ellipsoidal surfaces which satisfy the constraint that the total DM annihilation rate is 10 times the thermal rate.

We consider four different dark matter masses for which this constraint is satisfied, $m_{\chi}=10,75,150 \text{ and, } 200$ GeV.  For each DM mass, certain combinations of the operator coefficients will satisfy the rate constraint.   As the DM mass is decreased, lower cut-off scales $\Lambda$ will obey the annihilation rate constraint, just as in the previous section.  For each point on the ellipsoid, we may check if the combined photon flux from di-boson annihilations violates or satisfies the dwarf constraints.  Just as before, we find that parameter regions of low mass (hence low effective cut-off scale) are ruled out.  In our scan of Dark Matter masses, we found that there is no allowed parameter space for DM masses much below 10 GeV which also satisfies the constraint $\vev{\sigma v}_{\text{tot}} = 10 \vev{\sigma v}_{Th}$.

\begin{figure}[H]
\centering
\includegraphics[scale=0.35]{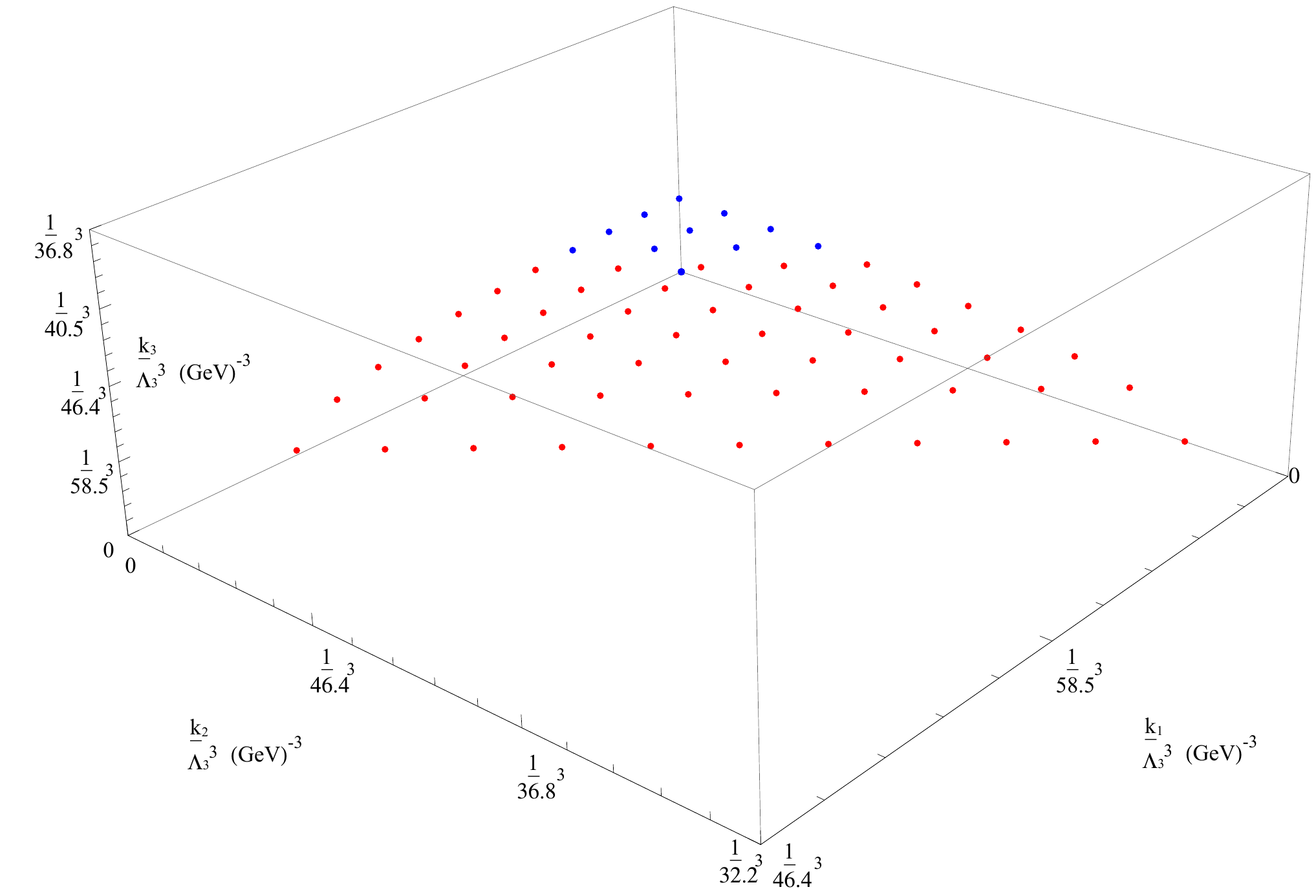}
\includegraphics[scale=0.40]{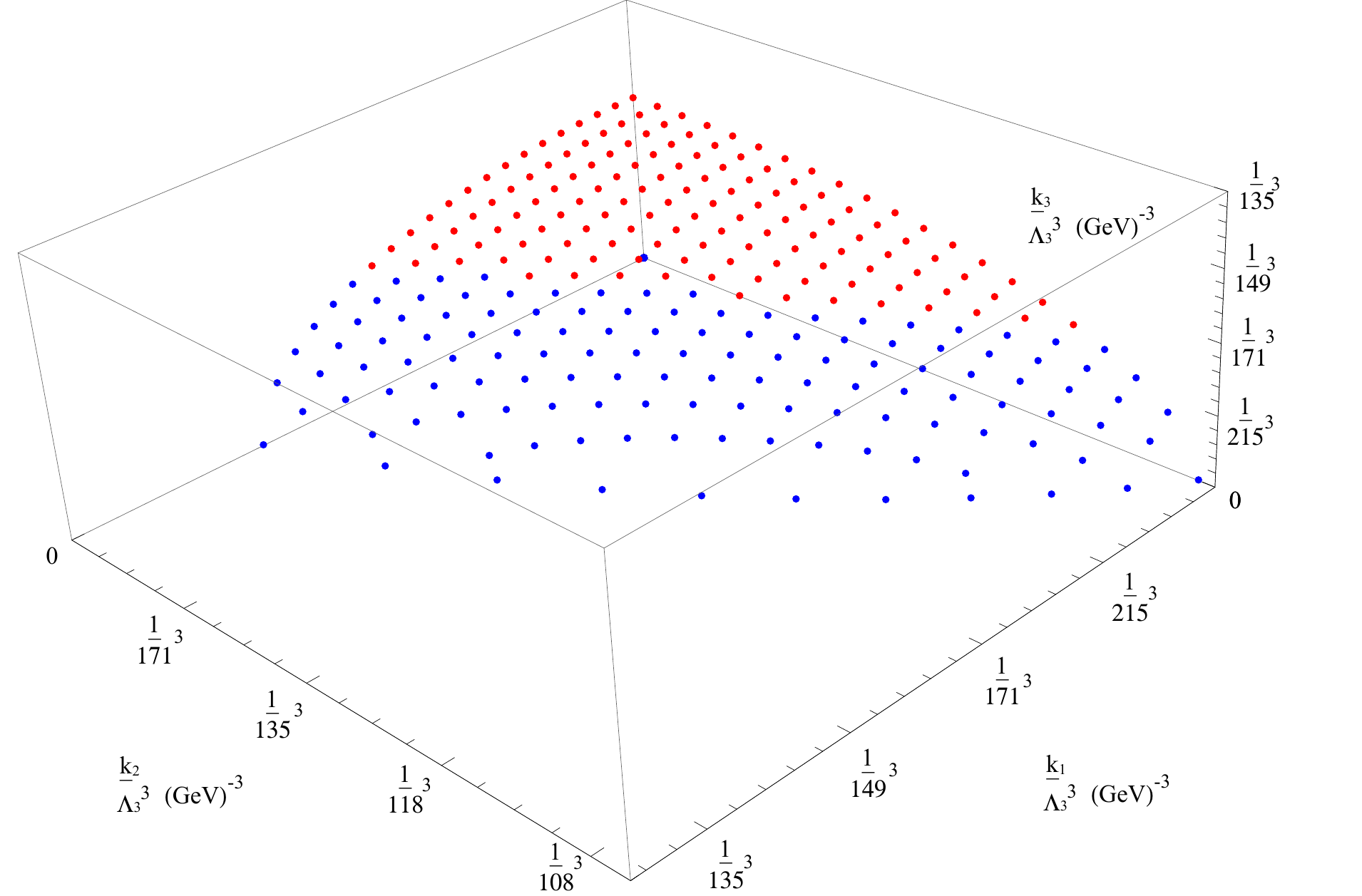}
\includegraphics[scale=0.40]{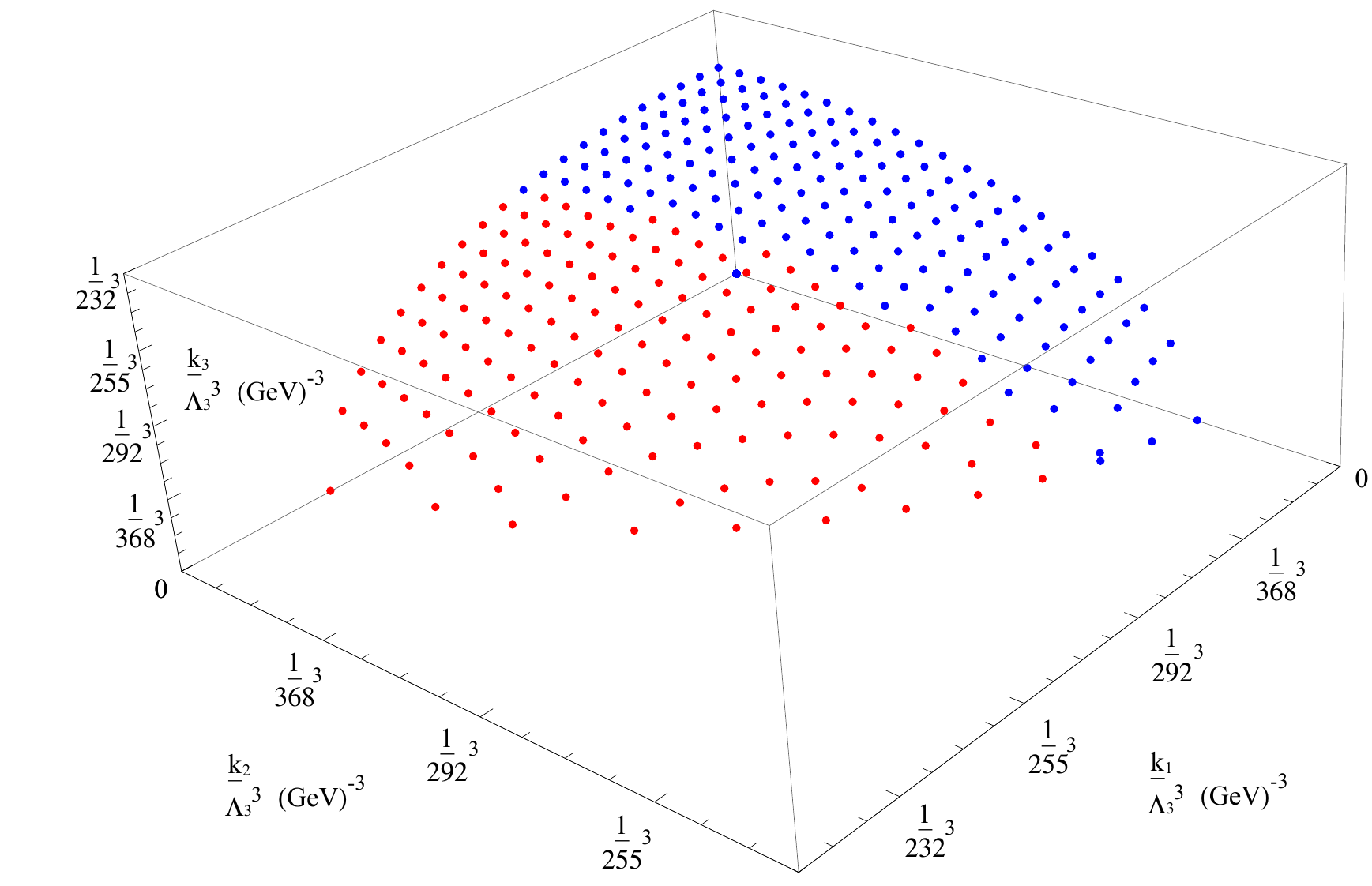}
\includegraphics[scale=0.35]{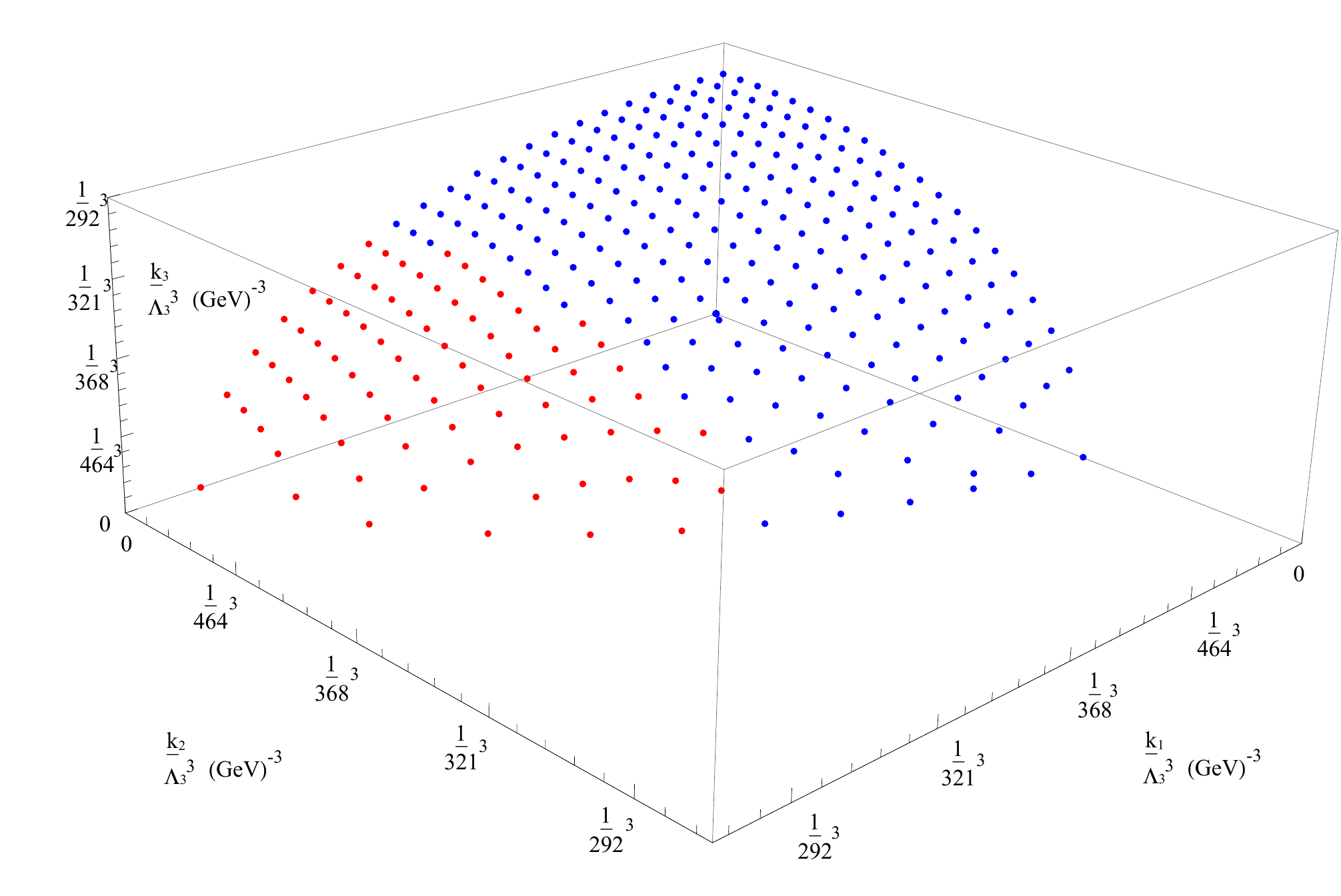}
\caption{Limits on effective operator coefficients for 10 times the thermal annihilation rate given by Fermi dwarf bounds. The axes show the value of $k_i/ \Lambda^3$.  Bounds on operators are shown for four dark matter masses, $m_{\chi}=10,75,150 \text{ and, } 200$ GeV (top left, top right, bottom left and bottom right respectively).  Red points indicate that the operator coefficient is ruled out, blue points are unconstrained.  }
\label{fig:yesno}
\end{figure}

In the  Fig. \ref{fig:yesno} plots, the red points are ruled out by the Fermi-LAT data, while the blue points are unconstrained.  As we scan through parameter space increasing the Dark Matter mass different annihilation channels open up which contribute to the total photon flux.  The first plot gives constraints for a 10 GeV dark matter particle.  At such low masses $m_{\chi} < m_Z/2$, only annihilations to gluons and photons are kinematically accessible. The photon energy $E_{\gamma}$ is equal to the DM mass $m_{\chi}$ and the operator coefficient values that yield a high di-photon annihilation rate and are ruled out.  For DM masses above half of the Z mass, the  $Z \gamma$ annihilation channel open up and starts to increase in magnitude but we see the the characteristic photon energy of the $Z\gamma$ annihilation is less than that of the photons in the $\gamma \gamma$ annihilation channel, here $E_{\gamma}= m_{\chi}\sqrt{1-4m_{\chi}^2/m_Z^2}$.  Therefore the electroweak operator coefficients in this mass region are not too strongly constrained.  For higher dark matter masses, $m_{\chi} > m_{W/Z}$, the WW and ZZ annihilation channels also open up as we see in the bottom two plots where $\m_{\chi}$ is 150 and 200 GeV respectively.

\begin{figure}[H]
\centering
 \includegraphics[width=.49\columnwidth]{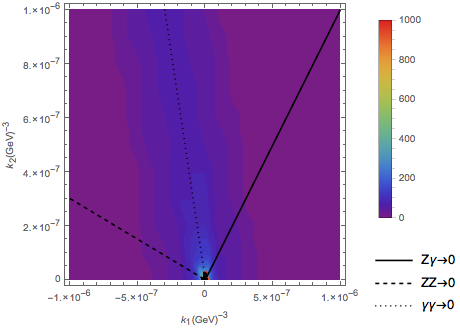}
 \includegraphics[width=.49\columnwidth]{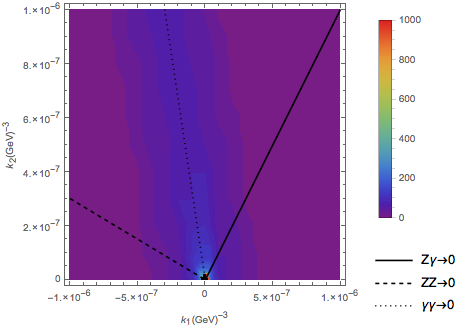}
 \includegraphics[width=.49\columnwidth]{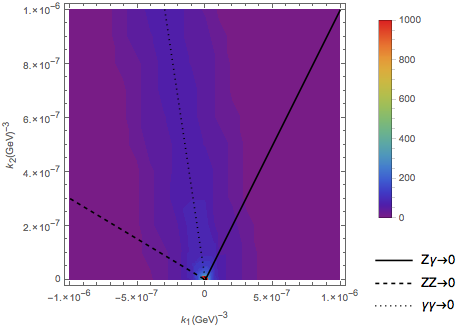}
\caption{These contours correspond to upper bounds on $m_{\chi}$ (in GeV) from annihilation into all final states allowed from the operators in Eq. \ref{eq:gaugelagr} with $k_{i}\equiv \kappa_{i}/\Lambda^{3}$.
The amount to the $gg$ final state is fixed in each of these plots as a percentage of the total rate, 0\% (upper left), 30\% (upper right), 70\% (lower).
The solid, dashed, and dotted lines correspond to a combination of $\kappa_1$ and $\kappa_2$ which turn off the final states $Z\gamma$, $ZZ$, and $\gamma \gamma$ respectively.
}
\label{fig:k1k2massbounds}
\end{figure}

\begin{figure}[H]
\centering
\includegraphics[width=.49\columnwidth]{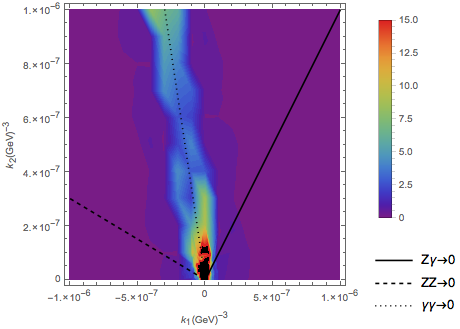}
\includegraphics[width=.49\columnwidth]{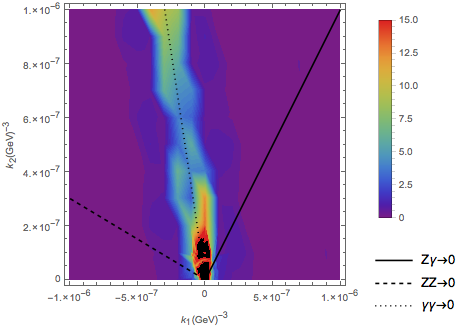}
\includegraphics[width=.49\columnwidth]{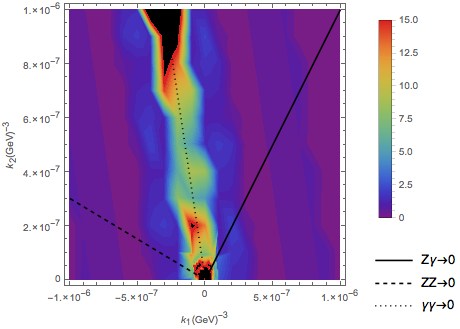}
\caption{These contours correspond to upper bounds on $\langle \sigma v \rangle_{\text{tot}}$ in units of the thermal relic, $\langle \sigma v \rangle_{\text{Th}}$, from annihilation into all final states allowed from the operators in Eq. \ref{eq:gaugelagr} with $k_{i}\equiv \kappa_{i}/\Lambda^{3}$.
The amount to the $gg$ final state is fixed in each of these plots as a percentage of the total rate, 0\% (upper left), 30\% (upper right), 70\% (lower).
The solid, dashed, and dotted lines correspond to a combination of $k_1$ and $k_2$ which turn off the final states $Z\gamma$, $ZZ$, and $\gamma \gamma$ respectively.
}
\label{fig:k1k2xsecbounds}
\end{figure}

We have so far proceeded by fixing the total annihilation rate.  We will now present the same constraints in a semi-orthogonal plane of parameter space where the total DM annihilation rate is allowed to vary.  In Figs. \ref{fig:k1k2massbounds} and \ref{fig:k1k2xsecbounds}, we show the Fermi-LAT bounds in plane of operator coefficients.  For simplicity in the figures we made the redefinition,  $k_i\equiv \kappa_{i}/\Lambda^{3}$.   We now allow the coefficients $k_{1}\equiv \kappa_{1}/\Lambda^{3}$ and $k_{2}\equiv \kappa_{i}/\Lambda^{3}$ to vary freely.  Here, $k_{3}$ is determined at each point by fixing the fractional annihilation rate to gluons $R_{gg}$.  We may then scan through increasing masses and determine which points in parameter space are ruled out. Note, we are not holding $k_{3}\equiv \kappa_{3}/\Lambda^{3}$ constant; as $m_{\chi}$ is varied this parameter must be varied in order for $R_{gg}$ to remain at a fixed value.  Each point in the plot corresponds to a unique value of the three operator coefficients so each specifies a particular total annihilation rate for DM.  As in the previous section, the parameter space exclusions are derived from the upper bounds on $\gamma$ ray flux from dwarfs.  For fixed operator coefficient values, the annihilation cross section goes as $m_{\chi}^4$, so the total photon flux for each point increases goes like $m_{\chi}^2$.  We may thus put an upper bound on DM masses for fixed values of operator coefficient.

We plot upper limits on Dark Matter masses in Fig. \ref{fig:k1k2massbounds}.  The top left, top right, and bottom plots in Figs. \ref{fig:k1k2massbounds} and \ref{fig:k1k2xsecbounds} correspond to $0\%$, $30\%$, and $70\%$ annihilation rate into di-gluons respectively.  The solid, dashed, and dotted lines represent the regions where $R_{Z\gamma}$, $R_{ZZ}$, and $R_{\gamma \gamma}$ go to zero respectively.  The vanishing of each of these channels coincide with a specific choice of the couplings in Eq. \ref{eq:gaugelagr} given by
\begin{eqnarray}
\kappa_1 = \kappa_2 \Rightarrow R_{Z\gamma} =0 \nonumber \\
\kappa_2 = -\kappa_1\frac{s_w^2}{c_w^2} \Rightarrow R_{ZZ} =0 \nonumber \\
\kappa_2 = -\kappa_1\frac{c_w^2}{s_w^2} \Rightarrow R_{\gamma \gamma} = 0
\end{eqnarray}
 As one would expect, the $\gamma\gamma$ channel tends to be the most powerful in constraining annihilation rates when allowed.
As we can see in Fig. \ref{fig:k1k2massbounds}, much of the parameter space will be ruled out by the $\gamma\gamma$ channel since the limits are stringent enough that other channels do not turn on before photon lines rule them out.

Notice that when the operator coefficients are very small, flux bounds allow for much larger DM masses.  If we look at the region where the $\gamma\gamma$ final state is off in parameter space, we still expect annihilations into the final states with $WW/ZZ$ and $Z\gamma$  which allow for strong constraints to be placed. In Fig. \ref{fig:k1k2xsecbounds} we plot the upper bound on the total DM annihilation rate in the plane of effective operator coefficients.

\subsection{Limits of the EFT and Collider Constraints}

We will now say a few words about the maximum range of validity of the EFT for DM annihilation processes. The most natural completion of a model with a vector boson portal involves loops of heavy messenger particles which are charged under the SM gauge groups such as those found such as in \cite{Weiner:2012gm}.  For the EFT to be valid, the momentum transfer of the process should be less than twice the messenger mass. Since the square root of the c.o.m energy of the process is twice the Dark Matter mass, we get $m_{\chi} < M_{\text{mess}}$.  We assume that the loops roughly scale like $f \alpha_i g_{\chi}^2 / M_{\text{mess}}^3$  where f is some numerical factor, $M_{\text{mess}}$ a messenger mass, and $g_\chi$ the coupling of the messenger to the dark matter.  Therefore, our effective operator coefficient $k_i /\Lambda^{3}$ corresponds to $f \alpha_i g_{\chi}^2/ M_{\text{mess}}^3$.  We can then estimate the messenger masses in terms of our effective cut-off for the maximum perturbative value of the hidden sector coupling $g_{\chi}$.  We find $M_{mess}\rightarrow \Lambda\left(f \alpha_i 16\pi^2 \right)^{-1/3}.$  For EFT validity $m_{\chi}< \Lambda\left(f \alpha_i 16\pi^2 \right)^{-1/3}$.  For order 1 values of f,  the validity limits for the U(1) operator for example are roughly  $m_{\chi} < \Lambda$; in regions of parameter space where the DM mass larger than the effective cut-off, derived bounds are not reliable.

We now wish to synthesize our parameter space constraints with EFT-validity limits and with constraints from colliders.  We will first reinterpret our results given in Fig. \ref{fig:k1k2xsecbounds} in the plane of dark matter mass vs operator suppression scale for several specific combinations of operator coefficients $\kappa_1$  $\kappa_2$ and $\kappa_3$.  First, we will assume $\kappa_3=0$, that is there are no annihilation of DM into gluons.  We now have a three dimensional parameter space  of $m_{\chi},k_1/\Lambda^{3}, k_2/\Lambda^{3} $ .  Our exclusions in this parameter space from Fermi-LAT data are shown in Fig. \ref{fig:scale} for certain simple ratios of couplings $\kappa_1$ and $\kappa_2$.  The excluded regions of parameter space lay under the colored lines.  The dot and dashed lines in Fig. \ref{fig:scale} correspond to curves of constant total annihilation rate.

\begin{figure}[h]
\centerline{\includegraphics[width=.8\columnwidth]{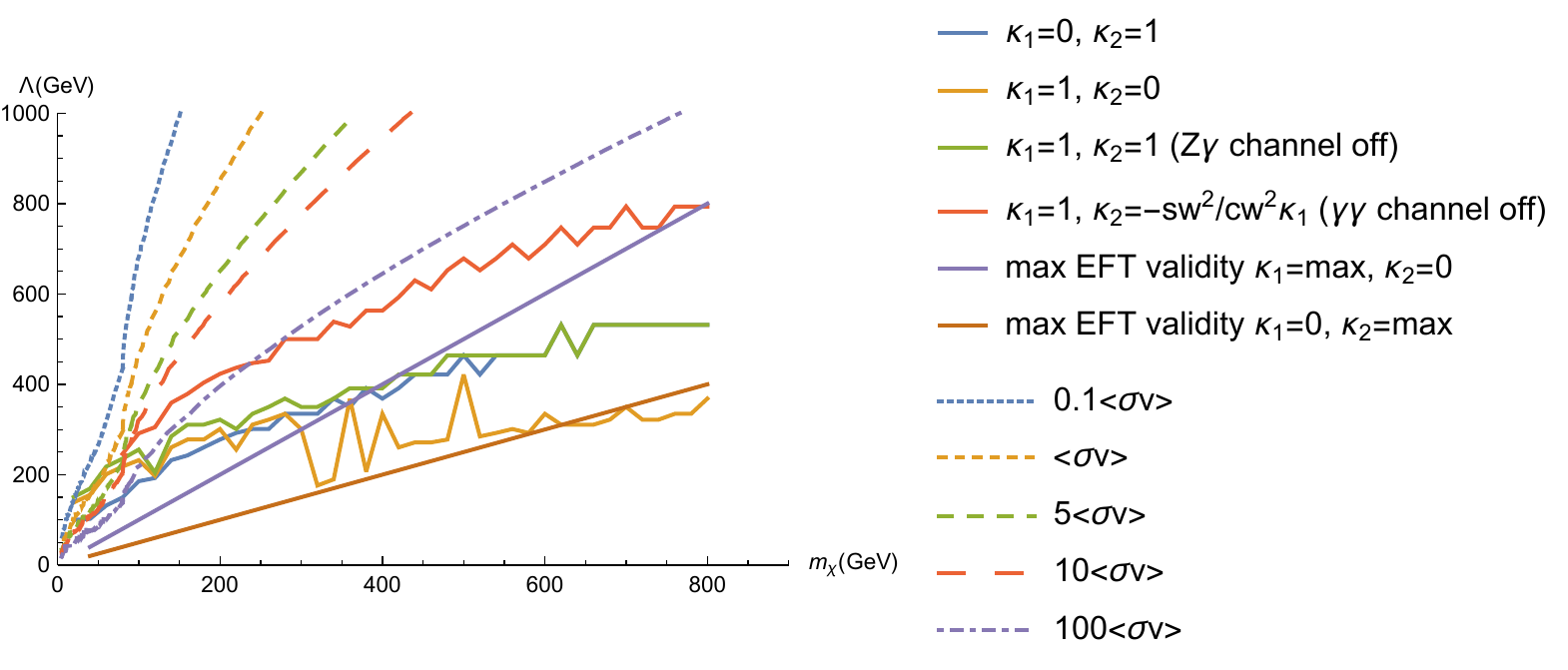}}
\caption{Constraints from Fig. \ref{fig:k1k2xsecbounds} reconfigured for the effective cut-off vs dark matter mass plane.  The regions below the solid lines are excluded.  The dashed lines correspond to curves of constant $\vev{\sigma v}_{\text{tot}}$}
\label{fig:scale}
\end{figure}

We recall that for the operators in Eq. \ref{eq:gaugelagr} $\vev{\sigma v} \propto m_{\chi}^4/\Lambda^6$ and thus upper bounds on annihilation cross section or dark matter mass correspond to lower bounds on the suppression scale $\Lambda$. Below the solid purple line is the region of breakdown of EFT validity for the indirect annihilation process for the U(1) operator.  Below the solid brown line is the region of breakdown of EFT validity for the indirect annihilation process for the SU(2) operator.

Vector boson portal models like those in this section have been widely studied in the context of collider production.  At LHC, operators of this type will lead to the production of dark matter particle pairs in association with a single vector boson, $pp \rightarrow \chi \overline{\chi}+V$ where V is a gluon, photon, W or Z.  No mono-boson signal is currently in excess and therefore collider constraints place lower bounds on the effective operator cut-offs $\Lambda_i$ for each given dark matter mass.   Studies on mono-boson collider signatures for the pseudo-scalar DM current may be found, for example, in references \cite{Lopez:2014qja} and \cite{Nelson:2013pqa}.  In the plot below we consider the effects of the SU(2) operator only for a simple comparison of collider and indirect detection bounds.  We show our exclusion contour from combined dwarf data, along with the ATLAS constrain from the mono-W channel (given by \cite{Lopez:2014qja}) for the pseudo-scalar current operator. We have also plotted contours of constant total DM annihilation rate.

\begin{figure}[h]
\centering
\includegraphics[width=.8\columnwidth]{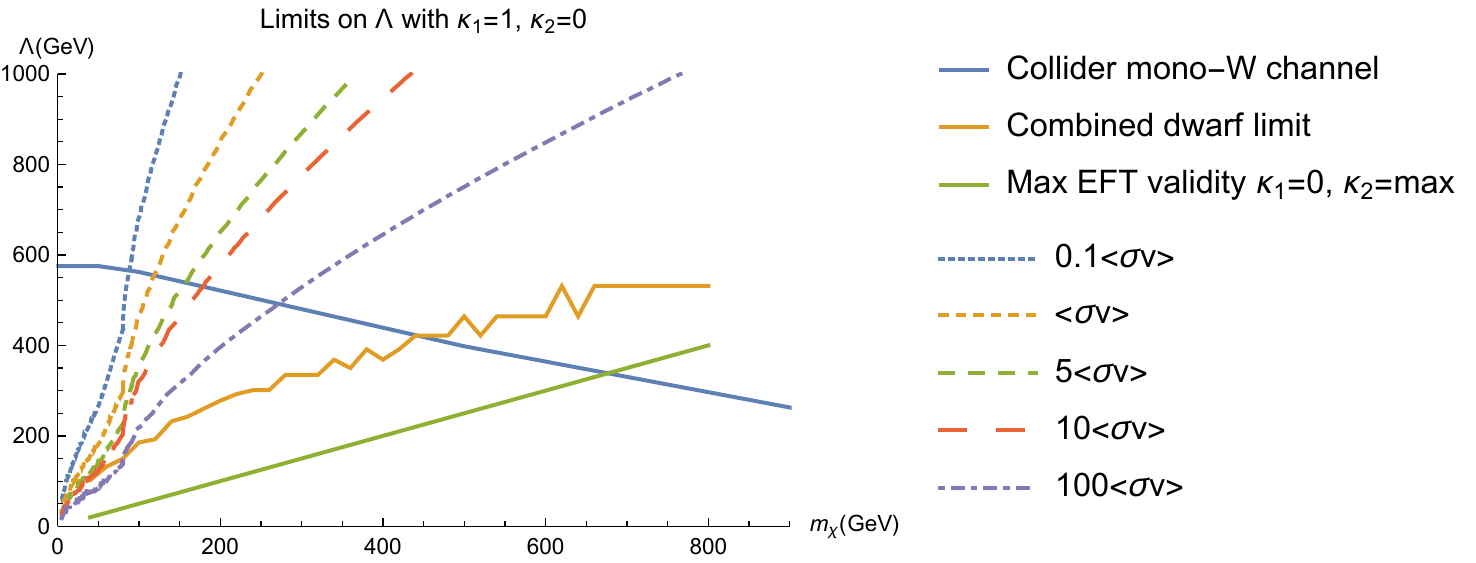}
\caption{Constraints from Fig. \ref{fig:k1k2xsecbounds} reconfigured for the effective cut-off vs Dark Matter mass plane}
\label{fig:scale}
\end{figure}

We see that following lines of constant annihilation rate we arrive at a minimum dark matter mass bound.  Following lines of constant operator coefficient we arrive at a maximum DM mass bound. The Fermi-LAT dwarf bounds are more constraining than collider analyses for DM masses above a few hundred GeV. It may be true that these bounds are more strict than collider bounds for regions of small $\Lambda$ as well. The validity of the EFT prescription for the collider bound requires that the momentum transfer $\sqrt{s} < 2 M_{\text{mess}} $.  The general procedure for collider bounds is to truncate the number of events, eliminating those where $\sqrt{s}$ is larger than 2 times the maximum messenger mass \cite{Busoni:2014sya}.  Previous collider analyses for models with D11 type gluon operators estimate that the region of maximum EFT validity limits interpretation for  cut-off scales below 350 GeV.  For these low effective cut-off scales, event rates are severely truncated \cite{Aad:2015zva}.  In our analysis for the SU(2) and U(1) operators, cut-off scales of around 500 GeV, imply messenger masses at or under roughly 1 TeV.  We generated a sample of mono-W events for the process $pp\rightarrow W \chi \overline{\chi}$ at 8 TeV c.o.m. energy using in MADGRAPH \cite{Maltoni:2002qb}. Barring any other cuts, we found that less than 10 percent of the events pass the above truncation requirement for an effective cut-off of 500 GeV and DM masses of 100 GeV.  We do not know how far under 500 GeV the collider limits hold for these operators and this is a matter for further investigation.

\section{Conclusions}
We have given indirect detection constraints derived from Fermi-LATs dwarf spheroidal galaxy data on variety of models with multiple final state annihilations of DM.  These models included dark matter portals from third generation fermions, SM vector bosons, and popular effective field theory models formally analyzed in collider physics.  We have presented constraints in model space on surfaces of fixed total annihilation rate, and in planes of freely varying effective operator coefficients.  In planes of fixed annihilation rate our bounds take the form of lower mass limits for dark matter mass/effective operator cut-off.  Our mass bounds are in the 10-100s of GeV range, and we have presented these results using a visualization on triangles and ellipsoidal sections.  We have also compared our constraints to those obtained by collider production for vector boson portal models.  In general we find that dwarf constraints overtake collider constraints for dark matter masses greater than a few hundred GeV, and in regions of low effective cut-off where there is theory break down for collider applications.  While existing results indicate that EFT validity fails under a few hundred GeV in collider studies, the exact range of maximum validity in vector boson portal models is yet to be determined. This is a course for future work.

Dwarf constraints used are among the most limiting constraints for Dark Matter models, and therefore present the best possibility of ruling out regions of Dark Matter parameter space.  We reiterate that a more sophisticated combination of dwarf data, which doesn't assume annihilation into only one channel would provide much tighter constraints on our parameter space, and this is another avenue for further work.  We also give another caveat here, we have implicitly assumed the most uncontroversial dark matter profiles for dwarf galaxies.  Assumptions of different Dark Matter profiles will lead to wiggle room in our constraints and require further study.

\section{Appendix}
We present lower limits on the effective cut-offs $\Lambda_b$ and $\Lambda_{\tau}$ for our third generation fermionic model.  We present lower limits for the fixed total DM annihilation rate $10\vev{\sigma v}_{Th}$.

\begin{figure}[H]
\centering
\includegraphics[scale=0.4]{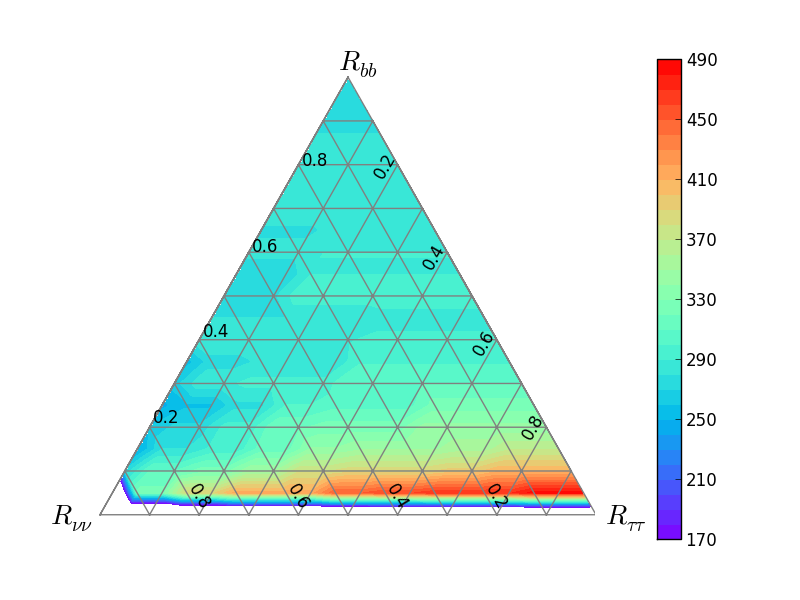}
\includegraphics[scale=0.4]{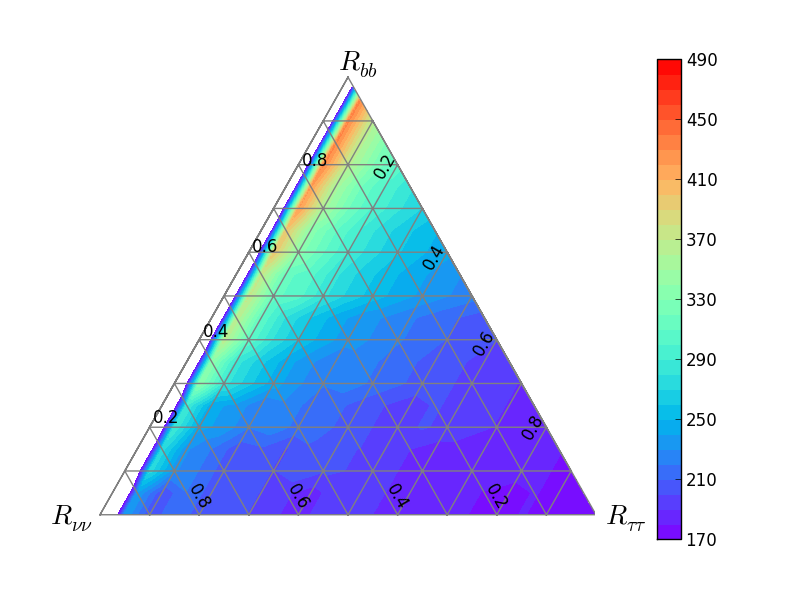}
\caption{Lower bound on operator coefficients $\Lambda_b^{*}$ and $\Lambda_{\tau}^{*}$}
\label{fig:LambdaLimits10Trel}
\end{figure}

\section{Acknowledgements}
This work was made possible with funds from DOE grant DE-SC0013529.  We would like to thank Archana Anandakrishnan, Kenny Ng, and John Beacom for input and conversations.

\end{document}